\documentclass[amsmath,floatfix,twocolumn,superscriptaddress]{revtex4-1}
\usepackage{amssymb,amsmath}
\usepackage{graphicx,array}
\usepackage{dcolumn}
\usepackage{psfrag}
\usepackage{bm}
\usepackage{color}

\begin{document}
\title{Universal ion-transport descriptors and classes of inorganic solid-state electrolytes}

\author{Cibr\'an L\'opez}
\affiliation{Departament de F\'isica, Universitat Polit\`ecnica de Catalunya, 08034 Barcelona, Spain}
\affiliation{Barcelona Research Center in Multiscale Science and Egineering, Universitat Politècnica de Catalunya, 08019 Barcelona, Spain}
\affiliation{Institut de Ci\`encia de Materials de Barcelona, ICMAB-CSIC, Campus UAB, 08193 Bellaterra, Spain}

\author{Agust\'i Emperador}
\affiliation{Departament de F\'isica, Universitat Polit\`ecnica de Catalunya, 08034 Barcelona, Spain}

\author{Edgardo Saucedo} 
\affiliation{Barcelona Research Center in Multiscale Science and Egineering, Universitat Politècnica de Catalunya, 08019 Barcelona, Spain}
\affiliation{Department of Electronic Engineering, Universitat Polit\`ecnica de Catalunya, 08034 Barcelona, Spain} 

\author{Riccardo Rurali}
\affiliation{Institut de Ci\`encia de Materials de Barcelona, ICMAB-CSIC, Campus UAB, 08193 Bellaterra, Spain}

\author{Claudio Cazorla}
\affiliation{Departament de F\'isica, Universitat Polit\`ecnica de Catalunya, 08034 Barcelona, Spain}
\affiliation{Barcelona Research Center in Multiscale Science and Egineering, Universitat Politècnica de Catalunya, 08019 Barcelona, Spain}

\maketitle

{\bf Solid-state electrolytes (SSE) with high ion conductivity are pivotal for the development and large-scale adoption 
of green-energy conversion and storage technologies such as fuel cells, electrocatalysts and solid-state batteries. Yet, 
SSE are extremely complex materials for which general rational design principles remain indeterminate. Here, we unite 
first-principles materials modelling, computational power and modern data analysis techniques to advance towards the solution 
of such a fundamental and technologically pressing problem. Our data-driven survey reveals that the correlations between ion 
diffusivity and other materials descriptors in general are monotonic, although not necessarily linear, and largest when the 
latter are of vibrational nature and explicitly incorporate anharmonic effects. Surprisingly, principal component and k-means 
clustering analysis show that elastic and vibrational descriptors, rather than the usual ones related to chemical composition 
and ion mobility, are best suited for reducing the high complexity of SSE and classifying them into universal classes. Our 
findings highlight the need of considering databases that incorporate temperature effects to improve our understanding of SSE 
and point towards a generalized approach to the design of energy materials.}
\\

Social networks use modern data analysis techniques to improve their customers experience and increase advertising 
revenues \cite{sumpter18}. Each mouse click and fingers action on the touchscreen reveal information on the users 
preferences that can be employed to classify individuals into similarity groups and thus better select the contents 
they are exposed to. Materials, in analogy to humans, conform to highly diverse and complex collectives and as such 
advanced data analysis techniques are being increasingly applied on them to improve their design and recommend 
possible uses \cite{kalinin15,tshitoyan19}. A necessary condition for the meaningful development and application 
of data-driven materials design strategies is the existence of comprehensive and reliable databases. 

Solid-state electrolytes (SSE) are a class of energy materials in which specific groups of ions may start to diffuse 
throughout the crystalline matrix driven by the thermal excitations \cite{hull04}. SSE are the pillars of green-energy 
conversion and storage technologies like fuel cells, electrocatalysts and solid-state batteries, hence tuning of their 
ion-transport properties turns out to be critical for the fields of Energy and Sustainability. SSE, however, are highly 
complex materials that present disparate compositions, structures, thermal behaviors and ion mobilities, thus it is 
difficult to ascribe them to general and rational design principles. These difficulties have motivated researchers to 
seek for easy to measure (or calculate) quantities that may serve as good descriptors of the ion conductivity; examples 
of such descriptors include structural parameters, defect formation energies, atomic polarizabilities and lattice dynamics 
\cite{guin15,bachman16,muy18,katcho19,muy20}. In recent years, pinpointing the role of phonon dynamics on ion transport 
has attracted special and increasing attention. Actually, for some specific SSE it has been demonstrated that lattice 
anharmonicity is one of the most influential factors affecting their ion mobility \cite{muy20,niedziela19,gupta22,cazorla19,gupta21,ding20}. 

Quantum mechanics-based density functional theory (DFT) \cite{cazorla17} has proven tremendously successful in the ﬁeld 
of computational materials science, and currently several databases of automated DFT calculations are being widely employed 
for materials design applications \cite{aflowlib,mp,oqmd,aiida}. Nevertheless, despite of their great successes, the existing 
DFT databases might not be entirely adequate for progressing in the design and understanding of SSE because they mostly contain 
information generated at zero temperature (e.g., structural parameters and formation energies) and thus completely disregard 
anharmonicity and $T$-induced effects \cite{kahle20}. In addition, modern high-throughput and machine learning studies relying 
on such DFT databases mainly have targeted Li and Na-based SSE families due to their predominance in electrochemical storage 
applications \cite{katcho19,zhang19,he19}. To holistically better understand the phenomena of ion transport, however, it might 
be necessary to analyse in equal measure other classes of SSE, like those involving mobile O, Cu, Ag and halide ions, which 
are technologically relevant as well \cite{hu21,aznar17,islam21}.

\begin{figure*}[t]
\centerline{
\includegraphics[width=1.00\linewidth]{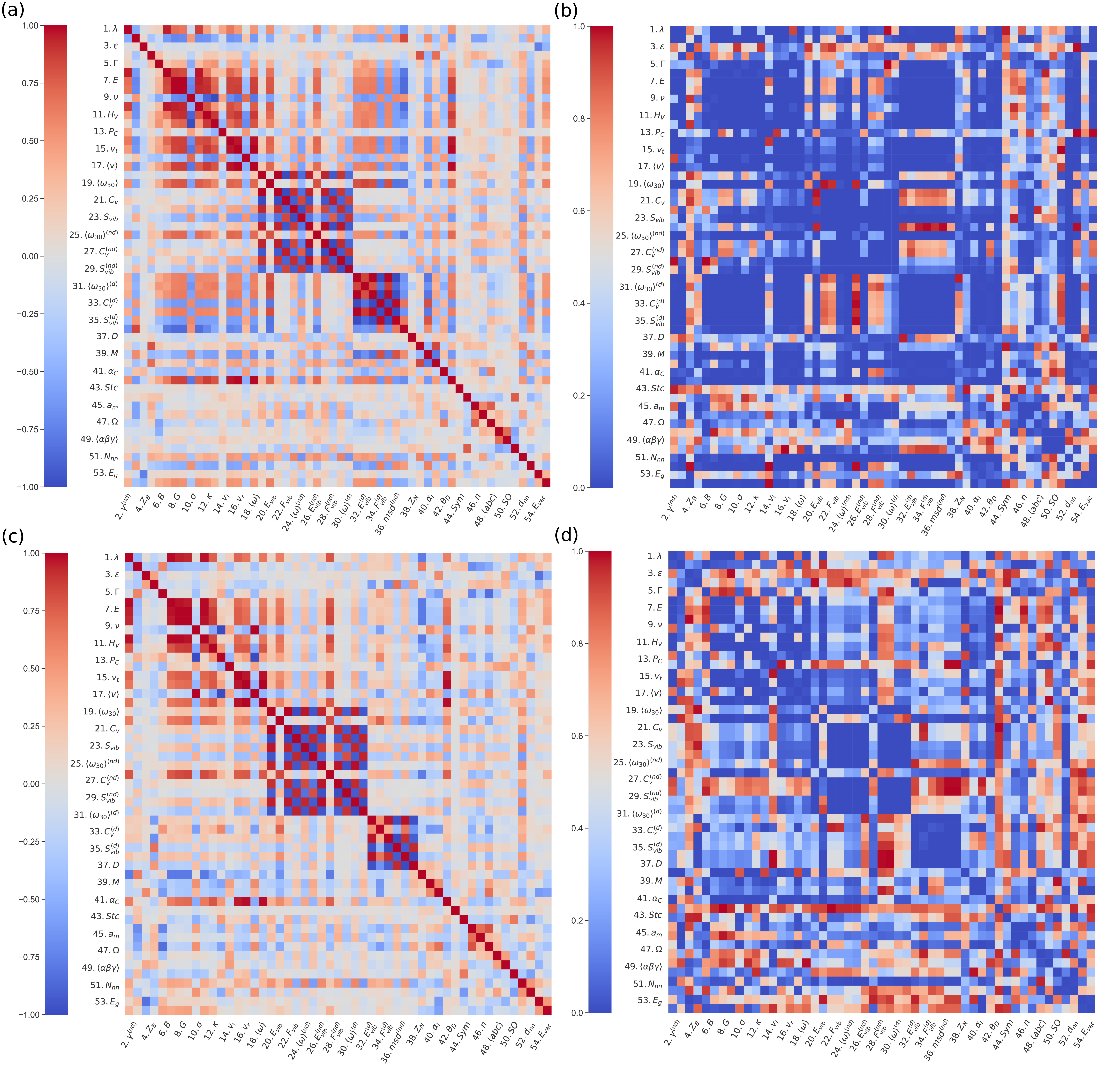}}
\caption{\textbf{Spearman correlograms and corresponding $p$-value matrices.} Correlations between pairs of materials
	features obtained for (a)~all and (c)~exclusively the Li-based SSE contained in our DFT-AIMD database. The $p$-value 
	matrices corresponding to all and exclusively Li-based Spearman correlograms are shown in (b) and (d), respectively.
	All the AIMD-based diffusive and vibrational descriptors were estimated at $T = 500 \pm 100$~K.}
\label{fig1}
\end{figure*}

Here, we present a data-driven analysis of SSE that covers aspects generally unaddressed by previous computational studies 
and the existing DFT materials databases. First, a comprehensive first-principles database was created for prototypical 
families of inorganic SSE containing both sets of zero-temperature DFT and finite-temperature \textit{ab initio} molecular 
dynamics (AIMD) results. Subsequently, a thorough correlation study between the ion diffusion coefficient ($D$) and other 
materials features was performed to determine universal ion-transport descriptors (as well as those specific to Li-based SSE). 
By relying on this new knowledge and the introduced DFT-AIMD database, several machine learning models were trained for the 
prediction of $D$ and other $T$-dependent quantities. Finally, principal component and k-means clustering analysis, data 
techniques customarily employed in the social sciences, were applied to reduce the high complexity of the SSE landscape and 
determine universal classes of fast-ion conductors.
\\

{\bf Curated first-principles SSE database.}~The generated SSE DFT-AIMD database \cite{database} comprises a total of $61$ 
materials of which $46$\% contain Li, $23$\% halide (i.e., F, Cl, Br and I), $15$\% Na, $8$\% O and $8$\% Ag/Cu atoms as the 
mobile ions. These percentages were selected in order to roughly reproduce the relative abundances of fast-ion conductors 
reported in the literature \cite{webofscience}. The generated SSE DFT-AIMD database contains materials with both stoichiometric 
and non-stoichiometric compositions and the AIMD results were obtained over a broad range of temperatures (Supplementary Tables 
1--3 and \cite{database}).     

To analyze the degree of similarity between all the surveyed SSE, a great variety of descriptors were estimated for each material 
adding up to a total of $54$ (the complete list of descriptors is detailed in the Methods section). Some of these descriptors had 
been already proposed in the literature (e.g., band gap and vacancy formation energy) while some others were totally new (e.g., 
harmonic phonon energy and Pugh's modulus ratio). The descriptors were classified into three general categories: ``mechanical-elastic'', 
``diffusive-vibrational'' and ``structural-compositional''. The value of some descriptors were obtained from zero-temperature DFT 
calculations (``mechanical-elastic'' and ``structural-compositional'') while the rest (``diffusive-vibrational'') were deduced 
from AIMD simulations performed at temperatures above ambient (Methods and Supplementary Tables 1--3). 

It is worth noting that the results obtained from AIMD simulations explicitly account for anharmonic effects, which constitutes 
one of the most important novelties and technical advances of the present work and introduced SSE database. Moreover, most 
vibrational descriptors were estimated considering the following cases (1)~all the ions, (2)~only non-diffusive ions and (3)~only 
diffusive ions, in order to better substantiate the role of the vibrating crystalline matrix on ion transport (Methods). The 
approximate computational cost of the generated SSE DFT-AIMD database was of $50$ Million CPU hours. 
\\

\begin{figure*}[t]
\centerline{
\includegraphics[width=1.00\linewidth]{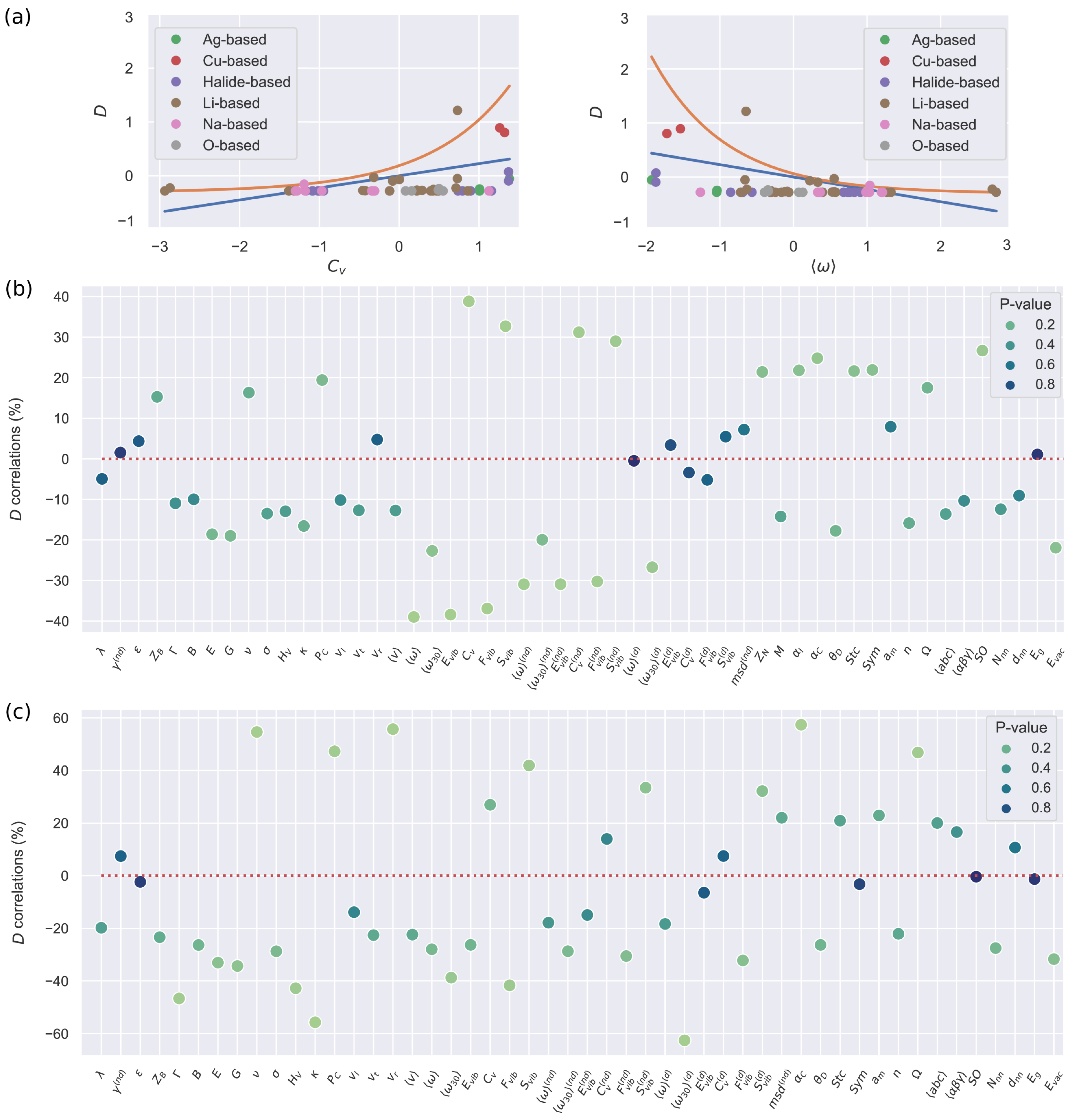}}
\caption{\textbf{Correlation study of the ion diffusion coefficient with other materials descriptors.} (a)~Standardized representation  
	 of the ion diffusion coefficient $D$ along with other materials descriptors. The descriptors correlations are, to some extent, 
	 monotonic but not linear as it is shown by the orange and blue lines therein (both simple guides to the eyes). Spearman correlation 
	 coefficients for $D$ and the rest of materials descriptors considered in this study, obtained by taking into account (b)~all and 
	 (c)~exclusively the Li-based compounds included in our DFT-AIMD database. The $p$-value results corresponding to the Spearman 
	 correlation coefficients are indicated with different colours. All the AIMD-based diffusive and vibrational descriptors were 
	 estimated at $T = 500 \pm 100$~K.}
\label{fig2}
\end{figure*}

{\bf Correlations between pairs of SSE descriptors.}~The correlation for a couple of materials descriptors, $x$ and $y$, can be 
quantified in several non-unique ways \cite{correlations}. In this work, we considered the Pearson ($c_{P}$) and Spearman ($c_{S}$) 
correlation coefficients which are defined like:  
\begin{eqnarray}
	c_{P}(x,y) & = & \frac{{\rm cov}(x,y)}{\sigma_{x} \sigma_{y}}~{\rm and} \nonumber \\
	c_{S}(x,y) & = & c_{P}\left[R(x), R(y)\right]~,
	\label{eq:corr}
\end{eqnarray}
where $\sigma_{i}$ is the standard deviation of the descriptor $i$ and $R(i)$ the rank of the $i$ samples. The covariance function 
is expressed as:
\begin{equation}
        {\rm cov}(x,y) = \langle xy \rangle - \langle x \rangle \langle y \rangle~,
	\label{eq:cov}
\end{equation}
where $\langle \cdot \rangle$ denotes expected value. The Spearman correlation coefficient is able to detect monotonic dependencies 
between pairs of descriptors while the Pearson can only identify linear correlations. Thus, the $c_{S}$ correlation coefficients 
are more general and robust than $c_{P}$ (i.e., can assess monotonic relationships whether linear or not). For this important reason, 
and despite the fact that linear correlations have been assumed in most previous SSE studies \cite{muy18,muy20}, we will stick to the 
Spearman correlation definition for the rest of our analysis.

Figure~\ref{fig1}a shows the Spearman correlation coefficients estimated for all pairs of materials descriptors considering 
the entire DFT-AIMD database (an analogous Pearson correlogram can be found in the Supplementary Fig.1). In view of the 
preeminence of Li-based SSE in electrochemical applications, the same correlation analysis was performed for this family of 
materials alone (Fig.\ref{fig1}c). To assess the statistical significance of the estimated $c_{S}$ correlograms, we computed 
the corresponding $p$-value matrices (Figs.\ref{fig1}b,d). The $p$-value represents the probability for a particular correlation 
result to arise if the null hypothesis (i.e., no correlation at all) were true, thus the smaller the calculated $p$-value the 
more statistically significant $c_{S}$ is. 

In a bird's eye view, the two correlograms obtained for all SSE and only those containing Li ions look quite similar. Nevertheless, 
the $p$-value matrix estimated for all SSE displays a noticeably higher number of statitiscally significant cases (arbitrarily 
defined here as $p < 0.2$), probably due to the larger amount of samples. Reassuringly, a number of already expected high correlation 
coefficients, like those estimated for couples of vibrational and elastic quantities that are physically related (e.g., $F_{vib}$ 
and $S_{vib}$), emerge from the calculated $c_{S}$ maps. For the sake of focus, hereafter we will concentrate on the correlations 
involving the ion diffusion coefficient ($D$). 

Figure~\ref{fig2}a encloses a standardized representation [that is, $\hat{x} \equiv \left( x - \langle x \rangle \right) / \sigma_{x}$] 
of the pairs of descriptors $D$--$C_{v}$ and $D$--$\langle \omega \rangle$, where $C_{v}$ stands for the lattice heat capacity and 
$\langle \omega \rangle$ for the average vibrational frequency (Methods). In these two cases, as well as in others not shown here, 
it is clearly appreciated that the dependency beween $D$ and the other quantities is far from linear although roughly monotonic. 
This outcome confirms that for determining reliable relationships between SSE features the Spearman correlation analysis is certainly 
more suitable than the usual Pearson approach. Actually, there are significant discrepancies between the Spearman and Pearson correlation 
maps; for instance, $c_{S}$ amounts to $-39$\% for the pair of descriptors $D$--$\langle \omega \rangle$ (Fig.\ref{fig1}a) whereas 
$c_{P}$ renders a significantly smaller value of $-23$\% (Supplementary Fig.2). 
\\

{\bf Universal ion diffusion descriptors.}~Figure~\ref{fig2}b shows the Spearman correlation coefficients estimated for all pairs 
of descriptors involving $D$ and considering the entire DFT-AIMD database. All the AIMD-based vibrational and diffusive descriptors
were estimated at $T = 500 \pm 100$~K. First, we note that larger $|c_{S}|$ values are associated with statistically more significant 
correlation results (i.e., smaller $p$-values). And second, the estimated correlation coefficients in general are not very high: only 
$19$ out of the $53$ pairs of materials descriptors present $|c_{S}|$'s larger than $20$\% while the maximum correlation value only 
amounts to $39$\% (obviously, the $D$--$D$ pair was excluded here). These low-correlation outcomes are consistent with the usual 
difficulties encountered in the settlement of flawless ion transport descriptors \cite{bachman16}.

Interestingly, the largest $D$ correlations are found for AIMD-based vibrational descriptors (Methods) like the phonon band center 
(or average lattice frequency), $\langle \omega \rangle$ ($-39$\%), lattice heat capacity, $C_{V}$ ($+39$\%), vibrational free energy, 
$F_{vib}$ ($-37$\%), and vibrational entropy, $S_{vib}$ ($+33$\%). These results indicate that insulator materials with small average 
phonon frequencies, large heat capacities and large vibrational entropies should be good ion conductors. It is worth noticing that 
strongly anharmonic materials perfectly fit into this description, thus our data-driven results generalize the conclusions of recent 
experimental SSE studies revealing that low-energy phonon modes can actively influence ion diffusion in some specific materials 
\cite{muy20,niedziela19,gupta22,cazorla19,gupta21,ding20}.

\begin{figure*}[t]
\centerline{
\includegraphics[width=1.00\linewidth]{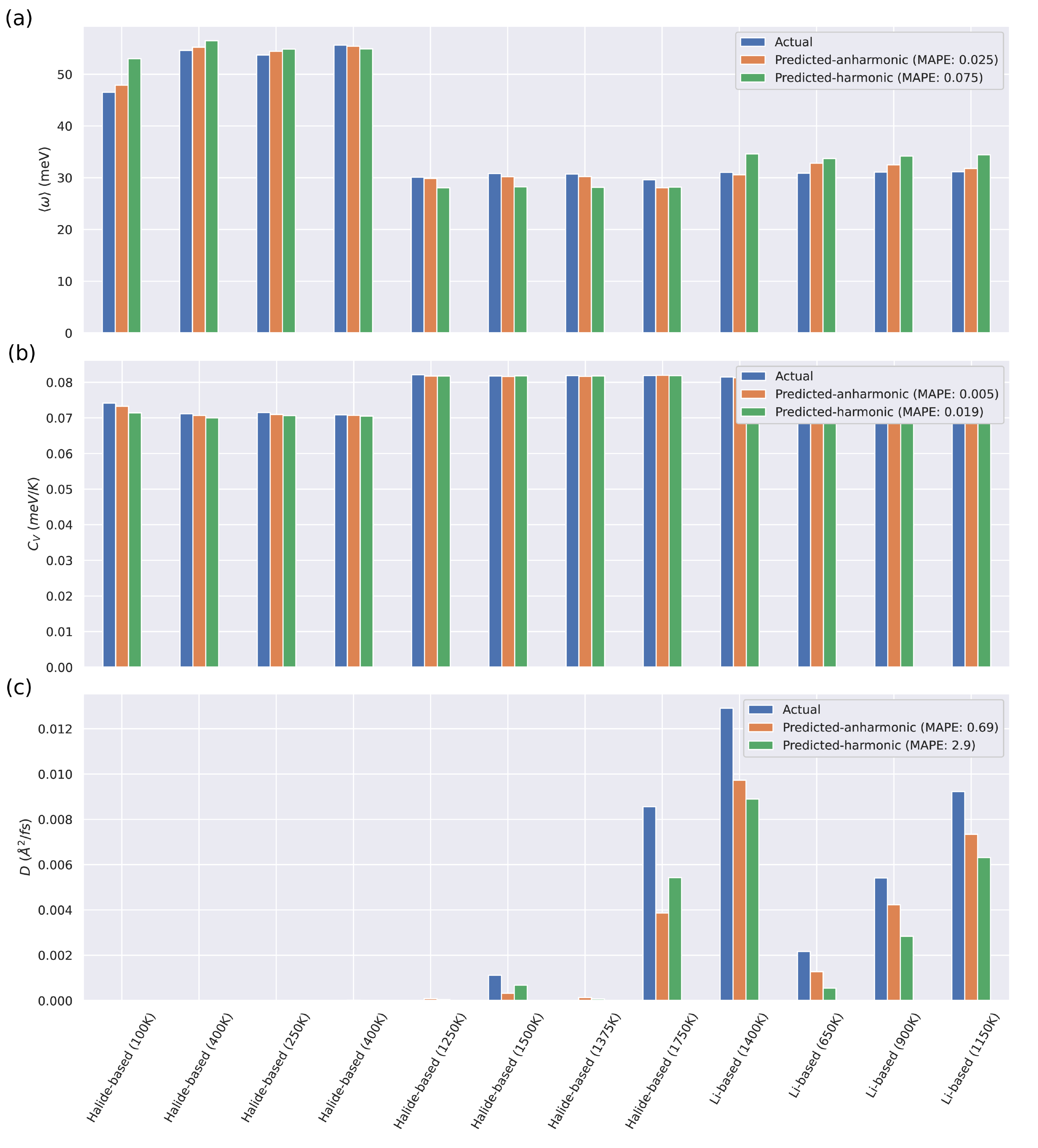}}
\caption{\textbf{Machine learning (ML) models trained in our DFT-AIMD database for prediction of different SSE $T$-dependent quantities.}
	The ML models were trained by considering and neglecting AIMD-based vibrational descriptors that explicitly incorporate anharmonic                      effects, labelled as ``anharmonic'' and ``harmonic'', respectively. (a)~First momentum of the vibrational density of states obtained 
        from AIMD simulations, $\left< \omega \right>$. (b)~Constant volume heat capacity obtained from AIMD simulations, $C_{V}$. (c)~Ionic
        diffusion coefficient obtained from AIMD simulations, $D$. ``MAPE'' stands for the mean absolute percentage error of the ML 
        predictions.
        }
\label{fig3}
\end{figure*}

Our correlation analysis provides further valuable insights. First, when the vibrational descriptors were estimated considering 
either non-diffusive or diffusive ions alone (superscripts ``nd'' and ``d'' in Fig.\ref{fig2}b, respectively) the value of the $D$ 
correlation coefficients slightly decreased in the first case ($|c_{S}| = 30$\%) and practically vanished in the second (except 
that corresponding to $\langle \omega_{30} \rangle^{(d)}$). This outcome highlights the existence of a strong and general interplay 
between the vibrating crystalline matrix and mobile ions. And second, when considering vibrational descriptors that do not explicitly 
take into account anharmonic effects, like the lowest-energy optical phonon mode calculated at $T = 0$~K ($\Gamma$ in Fig.\ref{fig2}b), 
the resulting $D$ correlation coefficient ($-11$\%) significantly drops in comparison to those obtained for anharmonic quantities 
(besides, the corresponding $p$-value increases). Thus, scrutinity of anharmonicity appears to be indispensable for the evaluation 
of reliable and statistically meaningful $D$ correlation coefficients.  

Few descriptors belonging to the ``structural-compositional'' category also correlate appreciably high with $D$. Of special mention are 
the vacancy formation energy of the mobile ions ($E_{vac}$, $-22$\%), the crystal polarizability ($\alpha_{C}$, $+25$\% --calculated with 
the Clausius-Mossotti relation--) and the symmetry of the perfect lattice ($SO$, $+27$\%) \cite{wang15}. On the other hand, intrinsically 
electronic properties like the energy band gap ($E_{g}$) and dielectric constant ($\epsilon$) have virtually no correlation with the ion 
diffusivity ($|c_{S}| \le 5$\%). As a word of caution, we note that when the correlations between $D$ and other materials descriptors are 
assumed to be linear (i.e., Pearson's approach) the resulting conclusions significantly differ from those just explained (Supplementary 
Fig.2). In particular, most $D$ correlation coefficients turn out to be smaller than the corresponding Spearman values and the materials 
descriptors belonging to the ``mechanical-elastic'' category (e.g., the Young and shear moduli --$E$ and $G$--) become equally relevant 
than the vibrational features. 

Figure~\ref{fig2}c shows the Spearman $D$ correlation coefficients estimated exclusively for Li-based SSE. Intriguingly, the resulting
$c_{S}$ chart differs appreciably from that estimated considering the entire DFT-AIMD database (Fig.\ref{fig2}b). First, the $D$ correlation 
coefficients in general present larger values with a total of $11$ pairs of materials descriptors scoring above $40$\%. Some of the largest 
$|c_{S}|$'s correspond to the AIMD-based vibrational descriptors $F_{vib}$ ($-42$\%), $S_{vib}$ ($+42$\%) and $\langle \omega_{30} \rangle^{(d)}$ 
($-63$\%). However, in contrast to the all-SSE case, now $\Gamma$, which is estimated at $T = 0$~K and does not explicitly account for 
anharmonicity, is strongly correlated with $D$ as well ($-47$\%). Moreover, several descriptors belonging to the ``mechanical-elastic'' 
category that, to the best of our knowledge, have not been previously proposed in the literature like the Vickers' hardness, $H_{V}$ 
($-43$\%), Pugh's modulus ratio, $\kappa$ ($-56$\%), Poisson's ratio, $\nu$ ($+55$\%), Cauchy's pressure, $P_{C}$ ($+48$\%), and velocity 
ratio, $v_{r}$ ($+56$\%), now also render very high $|c_{S}|$ values. Therefore, in terms of key $D$ descriptors, Li-based compounds are 
plainly different from the average SSE, a finding that fundamentally justifies the large number of studies focusing on the ion transport 
properties of this family of materials.   
\\

\begin{figure*}[t]
\centerline{
\includegraphics[width=1.00\linewidth]{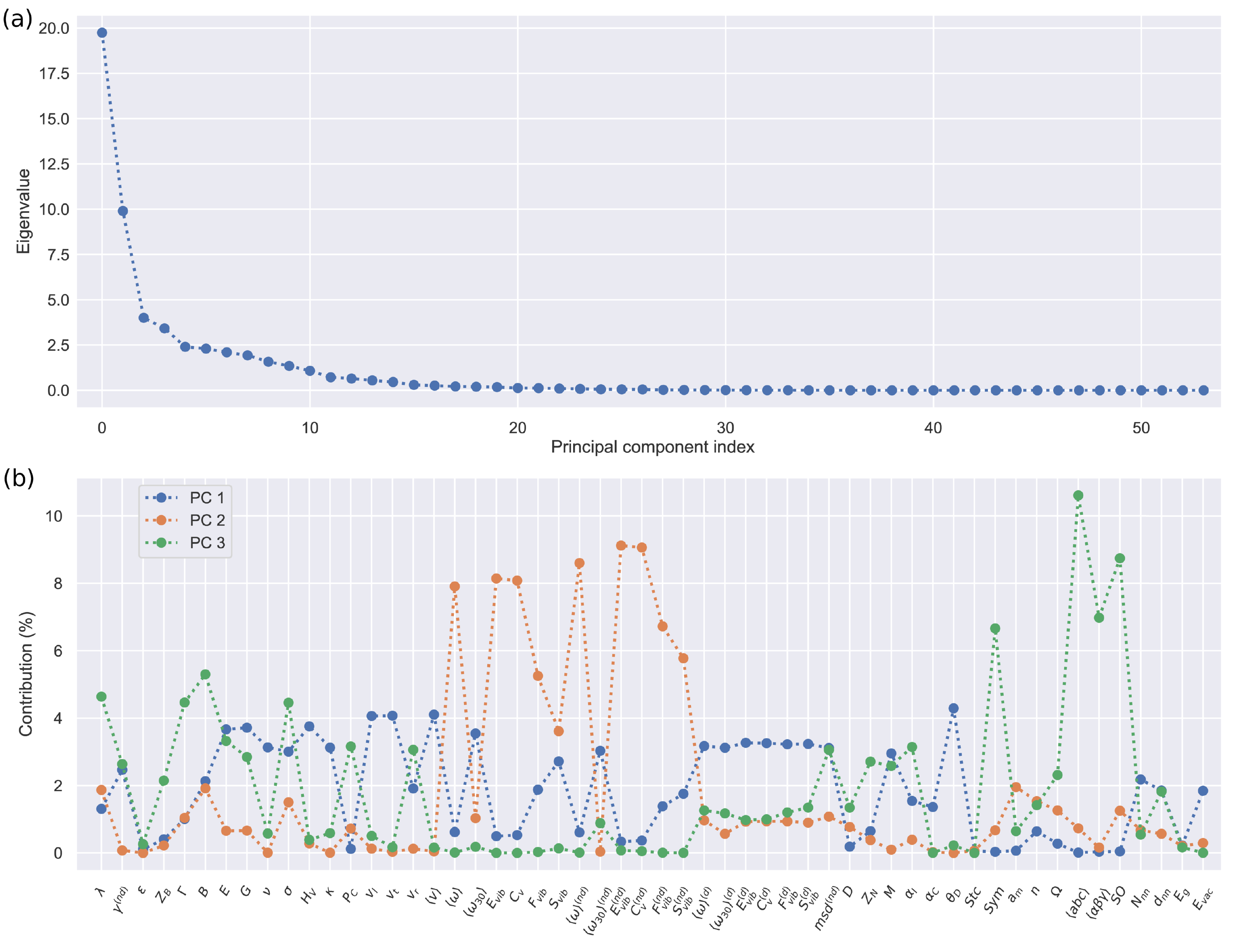}}
\caption{\textbf{Principal component analysis results obtained for the SSE DFT-AIMD database.} (a)~Eigenvalues corresponding
	to the diagonalization of the Spearman correlation matrix obtained by considering the entire DFT-AIMD database.  
	(b)~Eigenvector components of the first three principal components obtained from the diagonalization of the Spearman 
        correlation matrix obtained by considering the entire DFT-AIMD database.
	}
\label{fig4}
\end{figure*}

{\bf Machine learning models for prediction of $T$-dependent properties.}~In view of the complex relationships between $D$ and other 
materials descriptors (Fig.\ref{fig2}a), several machine learning (ML) models based on artificial neural networks were trained in our 
SSE DFT-AIMD database with the aim of predicting the ion diffusion coefficient and other relevant $T$-dependent properties of SSE such
as $\langle \omega \rangle$ and $C_{V}$ (Methods). We considered two different ML training schemes: (1)~considering all the materials 
descriptors (denoted as ``anharmonic'') and (2)~excluding the AIMD-based vibrational descriptors (``harmonic''). The predictions of our 
trained ML models for a validation set of $12$ compounds are shown in Fig.\ref{fig3}. Therein, it is appreciated that the two trained 
ML models can predict the finite-temperature values of $\langle \omega \rangle$ and $C_{V}$ with high accuracy. In particular, the mean 
absolute percentage error (MAPE) of the ``anharmonic'' (``harmonic'') ML model amounts to $2.5$\% ($7.5$\%) and only $0.5$\% ($1.9$\%) 
for $\langle \omega \rangle$ and $C_{V}$, respectively. In stark contrast, the ML predictions for the ion diffusion coefficient are much 
less accurate and there is a huge difference in the level of precision achieved with the ``anharmonic'' (MAPE of $69$\%) and ``harmonic'' 
($290$\%) ML models.  

Several conclusions follow from the ML results enclosed in Fig.\ref{fig3}. First, the SSE DFT-AIMD database introduced in this work 
appears to be comprehensive enough to ensure proper training of ML models able to make accurate predictions of certain $T$-dependent 
materials properties. And second, ML-based prediction of the ion diffusivity appears to be a particularly challenging task. In this latter
case, however, a big improvement is achieved when AIMD-based anharmonic vibrational descriptors are explicitly incorporated into the ML 
model (also in the $\langle \omega \rangle$ and $C_{V}$ cases). This outcome indirectly corroborates our previous finding that anharmonicity 
is a key general factor influencing ion transport. Nonetheless, to improve the ``anharmonic'' ML prediction of $D$ probably it is necessary 
to increase the number of SSE materials and descriptors in our DFT-AIMD database and/or resort to alternative and more advanced ML approaches 
(e.g., graph neural networks \cite{fung21}). 
\\

\begin{figure*}[t]
\centerline{
\includegraphics[width=1.00\linewidth]{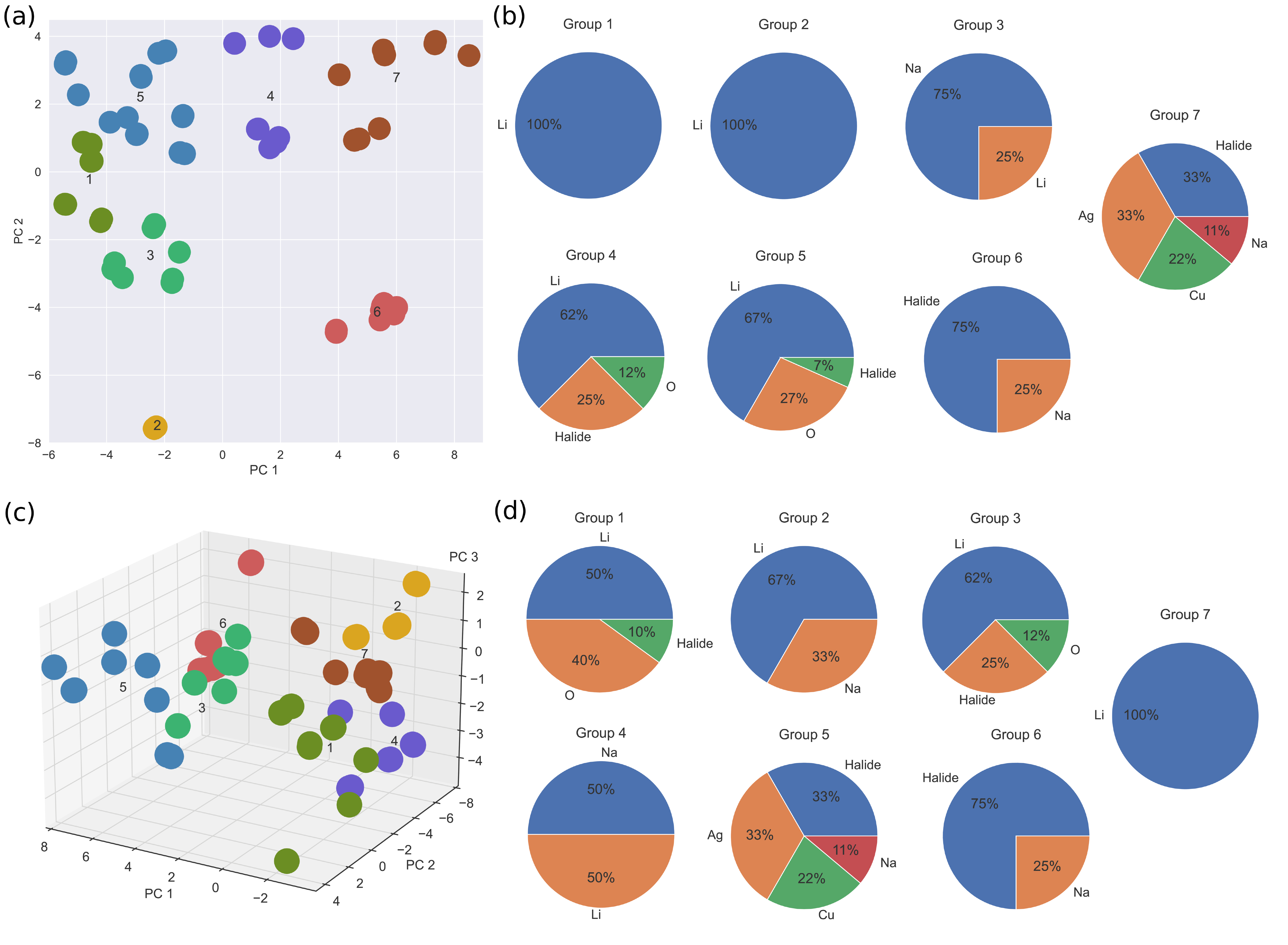}}
\caption{\textbf{K-means clustering analysis results obtained for the SSE DFT-AIMD database.} (a)~Classification of the analyzed
	materials in the orthogonal bidimensional space PC1--PC2. (b)~Materials population of each group identified in the PC1--PC2           
	space expressed in terms of the mobile ion species. (c)~Classification of the analyzed materials in the orthogonal tridimensional
	space PC1--PC2--PC3. (d)~Materials population of each group identified in the PC1--PC2--PC3 space expressed in terms of
	the mobile ion species. To improve visual clarity, some points have been removed from the plots without affecting the main
	conclusions.
        }
\label{fig5}
\end{figure*}

{\bf Complexity reduction in the SSE landscape.}~Principal component analysis (PCA) is a statistical technique widely employed for 
analyzing large datasets containing a high number of features. PCA increases the interpretability of a dataset by reducing its
dimensionality and simultaneously preserving the maximum amount of information. Complexity reduction is accomplished by linearly 
transforming the data into a new coordinate system where most of its variation can be described with fewer dimensions. The principal 
components are the eigenvectors of the dataset correlation matrix, which are expressed as linear combinations of the initial descriptors. 
The first principal component, the one with the largest eigenvalue, maximizes the variance of the projected data. The $i$-th principal 
component corresponds to a direction that is orthogonal to the previous $i-1$ principal components and along which the variance of the 
projected data is maximized as well. 

Figure~\ref{fig4} shows the results of diagonalizing the Spearman correlation matrix obtained for the entire SSE DFT-AIMD database. 
The first three principal components (PC) account for about two thirds of the total variance in the original $54$-dimensional dataset 
(as quantified by the sum of their normalized eigenvalues, $\approx 62$\%) hence its complexity can be greatly reduced by considering 
data projections on the orthogonal three-dimensional space PC1--PC2--PC3. PC1 presents a dominant ``mechanical-elastic'' character, 
PC2 ``vibrational'' and PC3 ``structural'' (Fig.\ref{fig4}b). Intriguingly, the contribution of the ion diffusivity to each of these 
PC's is practically zero, namely, $0.2$\% to PC1, $0.8$\% to PC2 and $1.3$\% to PC3. This data-driven outcome indicates that when it 
comes to characterize the great disparity of SSE, with the aim of fundamentally better understanding them and to establish general SSE 
categories, the ubiquitous $D$ descriptor is actually irrelevant. Likewise, the compound stoichiometry ($Stc$) and dielectric constant 
($\epsilon$) hardly contribute to the first three PC's hence they neither can be regarded as universally distinctive SSE features. By 
contrast, elastic and vibrational descriptors like $E$, $H_{V}$, $\langle \omega \rangle$ and $C_{V}$ become most pertinent for the 
evaluation of SSE similarities and general classification purposes.
\\

{\bf K-means clustering analysis.}~Figure~\ref{fig5} encloses the results of our k-means clustering analysis performed for the
entire SSE DFT-AIMD database. K-means clustering is an unsupervised learning algorithm that classify sets of objects in such a 
way that objects within the same group, called ``cluster'', are more similar to each other in a broad sense than to the objects 
in other clusters. We selected a subminimal number of $7$ clusters to account for the SSE database variance based on the outcomes 
of the elbow and silhouette methods (Supplementary Figs.3--4). This number of clusters is already larger than the number of 
$A$-based SSE families considered in this study (i.e., $6$ with $A =$ Li, Na, halide, Ag, Cu and O). Thus, it straightforwardly 
follows that the materials composition, despite of its obvious utility in naming compounds, should not be regarded as a fine 
descriptor of SSE diversity since, at least, one SSE family will spread over more than one k-means cluster. 

Figures~\ref{fig5}a--b show the results of our k-means clustering analysis performed in the simplified PC1--PC2 space. It is noted
that Li-based SSE are present in $5$ out of the total $7$ clusters. From those $5$ clusters, Li-based SSE are the most abundant in 
$80$\% of the cases and overall they share similarities with other Na, halide and O-based SSE (although not necessarily in terms of 
ion conductivity). In clusters number $1$ and $2$, which are respectively characterized by dominant PC1 (``elastic'') and PC2 (``vibrational'') 
components, Li-based SSE actually conform the entire population. From these outcomes, we may readily conclude that (1)~Li-based SSE  
are intrinsically different from Ag- and Cu-based SSE, which in turn are highly similar because inhabit the same cluster, and (2)~Li-based 
SSE can be partitioned into several similarity subgroups attending to their elastic and vibrational properties. Likewise, halide-based 
SSE appear in $4$ different clusters, Na-based in $3$ and O-based in $2$. Thus, as it was mentioned above, chemical composition is not 
a good descriptor for grouping SSE into similarity categories. 

Figures~\ref{fig5}c--d enclose the k-means clustering results obtained in the expanded PC1--PC2--PC3 space. In this case, the main 
findings are very similar to those just explained for the reduced P1--P2 space, namely, Li-based SSE are present in $5$ out of the 
total $7$ clusters and they are particularly numerous in the majority of those groups (e.g., $100$\% in cluster number $7$ and $67$\% 
in cluster number $2$). Likewise, halide-based SSE spread over $4$ different clusters, Na-based over $4$, O-based over $2$ and Cu/Ag-based 
only appear in $1$. The Li-based SSE family overall shares similarities with other Na, halide and O-based SSE (not so with Cu- and Ag-based 
SSE), and most subgroup differences (i.e., relative distances between clusters centroids) are contained in the P1--P2 plane. Thus, the 
PC3 (``structural'') dimension does not appear to add sensible information on SSE diversity and for grouping purposes is practically 
expendable (in accordance with its relatively small eigenvalue of $\approx 4$\%, Fig.\ref{fig4}a).

The presented k-means clustering analysis enlightens the difficulties encountered in the rational design of SSE with specific ion 
mobility. The bulk of the variation in the SSE family is encoded in the materials elastic and vibrational properties, neither 
in the ion mobility nor their chemical composition. This finding implies that materials which can be rigorously considered as 
overall highly similar (because they belong to a same k-means cluster) in practice may exhibit very different ion diffusion and 
chemical features (e.g., Li-based and halide-based SSE). Conversely, materials which render very similar ion mobilities and 
chemical compositions (e.g., Li-based SSE inhabiting groups $7$ and $3$ in Fig.\ref{fig5}d) may behave radically different in terms 
of other measurable quantities. These conclusions are consistent with the $D$ correlation results enclosed in Fig.\ref{fig2}, which 
show that Li-based SSE can significantly depart from the general trends averaged over all SSE.  
\\

In summary, we have presented an original and comprehensive SSE data-driven study on the correlations of the ion diffusion with 
other materials descriptors as well as a rigorous examination of universal SSE categories, based on a new and thorough DFT-AIMD 
database comprising both zero-temperature and finite-$T$ first-principles results. It has been demonstrated that ion diffusion 
correlates strongly and monotonically, not necessarily linearly, with vibrational descriptors that explicitly incorporate anharmonic 
effects (i.e., are estimated from AIMD simulations). In the particular case of Li-based SSE, the ion mobility also correlates 
significantly with elastic quantities like the Vickers' hardness, Pugh's modulus ratio, Poisson's ratio and Cauchy's pressure, 
pertinent ion-diffusion descriptors that previously have been overlooked in the literature. Furthermore, most of the variation 
in the generated SSE $54$-fold dimensional space can be resolved in terms of elastic and vibrational descriptors; ion mobility 
and chemical composition are very much irrelevant when it comes to quantify the SSE diversity, a fact that complicates the rational 
design of SSE with targeted ion conductivities. The present data-driven study highlights the necessity to consider finite-temperature 
effects in a high-throughput fashion to better understand SSE and improve the predictions of machine learning models in them; it 
also provides new theoretical guidelines for analyzing materials that in analogy to SSE are highly anharmonic and technologically 
relevant (e.g., thermoelectrics and superconductors).  
\\

\section*{Methods}
\label{sec:methods}
{\bf First-principles calculations outline.}~\textit{Ab initio} calculations based on density functional theory (DFT) were 
performed to analyse the physico-chemical properties of bulk SSE. We performed these calculations with the VASP code \cite{vasp} 
by following the generalized gradient approximation to the exchange-correlation energy due to Perdew \emph{et al.} \cite{pbe96}. 
(For some halide compounds, possible dispersion interactions were captured with the D3 correction scheme developed by Grimme and 
co-workers \cite{grimmed3}.) The projector augmented-wave method was used to represent the ionic cores \cite{bloch94} and for each 
element the maximum possible number of valence electronic states was considered. Wave functions were represented in a plane-wave 
basis typically truncated at $750$~eV. By using these parameters and dense ${\bf k}$-point grids for Brillouin zone integration, 
the resulting zero-temperature energies were converged to within $1$~meV per formula unit. In the geometry relaxations, a tolerance 
of $0.005$~eV$\cdot$\AA$^{-1}$ was imposed in the atomic forces.
\\

{\bf First-principles molecular dynamics simulations.}~\emph{Ab initio} molecular dynamics (AIMD) simulations based on DFT 
were performed in the canonical $(N,V,T)$ ensemble (i.e., constant number of particles, volume, and temperature) for all the 
considered bulk materials. The selected volumes were those determined at zero temperature hence thermal expansion effects were
neglected; nevertheless, based on previously reported molecular dynamics tests \cite{cazorla19}, thermal expansion effects are 
not expected to affect significantly the estimation of the ion-transport properties of SSE at moderate temperatures (i.e., 
$T = 500 \pm 100$~K). The concentration of ion vacancies in the non-stoichiometric compounds was also considered independent of the 
temperature and equal to $\sim 1$--$2$\%. The temperature in the AIMD simulations was kept fluctuating around a set-point value 
by using Nose-Hoover thermostats. Large simulation boxes containing $N_{ion} \sim 200$--$300$ atoms were employed in all the cases 
and periodic boundary conditions were applied along the three Cartesian directions. Newton's equations of motion were integrated 
by using the customary Verlet's algorithm and a time-step length of $\delta t = 1.5 \cdot 10^{-3}$~ps. $\Gamma$-point sampling 
for integration within the first Brillouin zone was employed in all the AIMD simulations. The finite-temperature simulations 
typically comprised long simulation times of $t_{total} \sim 100$--$200$~ps. For each material, we ran an average of $3$ AIMD 
simulations at different temperatures and considering both stoichiometric and non-stoichiometric compositions (Supplementary
Tables 1--3 and \cite{database}). Previous tests performed on the numerical bias stemming from the finite size of the simulation cell and duration 
of the molecular dynamics runs reported in work \cite{cazorla19} indicate that the adopted $N_{ion}$ and $t_{total}$ values 
should provide reasonably well converged results for the ion diffusivity and vibrational density of states of SSE.
\\

{\bf Estimation of key diffusive and vibrational properties.}~The mean-squared displacement (MSD) was estimated like:
\begin{eqnarray}
{\rm MSD}(\tau) & = & \frac{1}{N_{ion} \left( N_{step} - n_{\tau} \right)} \times \\ \nonumber
                &   & \sum_{i=1}^{N_{ion}} \sum_{j=1}^{N_{step} - n_{\tau}} | {\bf r}_{i} (t_{j} + \tau) - {\bf r}_{i} (t_{j}) |^{2}~, 
\label{eq1}
\end{eqnarray}
where ${\bf r}_{i}(t_{j})$ is the position of the migrating ion $i$ at time $t_{j}$ ($= j \cdot \delta t$), $\tau$ 
represents a lag time, $n_{\tau} = \tau / \delta t$, $N_{ion}$ is the total number of mobile ions, and $N_{step}$
the total number of time steps. The maximum $n_{\tau}$ was chosen equal to $N_{step}/2$ in order to accumulate enough 
statistics to reduce significantly the fluctuations in ${\rm MSD}(\tau)$ at large $\tau$'s. The diffusion coefficient 
then was obtained by using the Einstein relation: 
\begin{equation}
D =  \lim_{\tau \to \infty} \frac{{\rm MSD}(\tau)}{6\tau}~.  
\label{eq2}	
\end{equation}
In practice, we performed linear fits over the averaged ${\rm MSD}(\tau)$ values calculated within the lag time interval 
$\tau_{max}/2 \le \tau \le \tau_{max}$.   

To estimate the vibrational density of states (VDOS) of bulk SSE considering anharmonic effects, $g(\omega)$, we calculated 
the Fourier transform of the corresponding velocity-velocity autocorrelation function as obtained directly from the AIMD 
simulations, namely:
\begin{equation}
	g(\omega) = \frac{1}{N_{ion}} \sum_{i}^{N_{ion}} \int_{0}^{\infty} 
	\langle {\bf v}_{i}(\tau)\cdot{\bf v}_{i}(0)\rangle e^{i\omega \tau} d\tau~,
\label{eq5}	
\end{equation}
where ${\bf v}_{i}(t)$ represents the velocity of the $i^{\rm th}$ atom at time $t$, and $\langle \cdots \rangle$
denotes statistical average in the $(N,V,T)$ ensemble. Once the density of vibrational states was determined, it was 
straightforward to calculate the corresponding phonon band center (or average lattice frequency), $\langle \omega \rangle$, 
defined like:
\begin{equation}
	\langle \omega \rangle = \frac{\int_{0}^{\infty} \omega~ g(\omega)~ d\omega}{\int_{0}^{\infty} g(\omega)~ d\omega}~,
\label{eq6}
\end{equation}
which also depends on $T$. Likewise, the contribution of a particular group of ions to the full VDOS was estimated by 
considering those ions alone in the summation appearing in Eq.(\ref{eq5}). In order to determine a characteristic low-energy 
phonon frequency for bulk SSE, we defined the quantity:
\begin{equation}
	\langle \omega_{30} \rangle = \frac{\int_{0}^{\omega_{max}} \omega~ g(\omega)~ d\omega}{\int_{0}^{\omega_{max}} g(\omega)~ d\omega}~,
\label{eq7}
\end{equation}
for which we imposed an arbitrary cut-off frequency of $\omega_{max} = 30$~meV. The analytical expression for other vibrational 
descriptors (e.g., $F_{vib}$, $E_{vib}$ and $C_{V}$) can be found in work \cite{phonopy}. 
\\

{\bf Machine learning models.}~The Scikit-learn package in Python \cite{scikit} was used to encode the non-numeric descriptors as 
well as to implement the Artificial Neural Network (ANN) conforming our machine learning model. For the generation of the input data, 
the simulations involving all compounds, compositions and temperatures in our SSE DFT-AIMD database were taken into consideration 
(i.e., a total of $174$ samples, Supplementary Tables 1--3 and \cite{database}). The non-numeric descriptors (i.e., the diffusive 
chemical element, stoichiometricity, chemical composition of the compound and symmetry of the relaxed structure) were encoded with 
the one-hot encoding approach, and all input data was normalized using a standard scaler. Specifically, a Multi-Layer Perceptron 
Regressor (MLPR) was implemented, consisting on input, hidden and output layers. As output layer, the algorithm was defined in such 
a way that any of the considered descriptors could be used as dependent variable. Consequently, the input layer was constructed as 
the set of all the other descriptors. Optionally, anharmonic descriptors could be removed from the input layer if desired. Finally, 
$6$ hidden layers of $150$, $500$, $50$, $150$, $70$ and $100$ neurons, respectively, showed the best performance. 

Attending to the extraction of metrics, K-fold validation was implemented: on each interation, the model was required to predict the 
output for one element using the rest as training set. Therefore, given that each element consists of a different number of simulations 
(the original dataset presents a variable number of simulated temperatures and stoichiometricities for each element), the computed 
metrics were weighted with the number of predicted outputs and then divided by the total amount of simulations. The optimization of 
the model was monitored by using the mean absolute percentage error (MAPE) defined like:
\begin{equation}
    MAPE = \frac{1}{N} \sum_{i=1}^N \left| \frac{x^0_i - x_i}{x^0_i} \right|~,
    \label{eq:mae}
\end{equation}
where $N$ is the total number of samples in the set, $\{x\}$ the predicted outputs and $\{x^0\}$ the actual values in the DFT-AIMD
database. Note that these metrics can be extracted from both the training and test sets. As optimal hyperparameters, Adam optimizer 
with the square error as loss function and constant learning rate of $0.001$, rectified linear unit (ReLU) activation function, and 
$\alpha = 0.05$ strength for the $L^2$ regularization term of the loss function were used.
\\

\begin{table*}[t]
    \begin{tabular}{ccc}
        \begin{tabular}{ccc}
\hline
\hline
		$$ & $$ & $$ \\
		Symbol              & Descriptor (M-E)     &    Estimation approach    \\ 
		$$ & $$ & $$ \\
\hline
\hline
		$$ & $$ & $$ \\
	    $\lambda$           & $1^{st}$ Lam\'e parameter                        & DFT         \\
            $B$                 & Bulk modulus                                     & DFT         \\
            $E$                 & Young modulus                                    & DFT         \\
            $G$                 & Shear modulus                                    & DFT         \\
            $\nu$               & Poisson's ratio                                  & DFT         \\
            $\sigma$            & P-wave modulus                                   & DFT         \\
            $H_V$               & Vickers' hardness                                & DFT         \\
            $\kappa$            & Pugh's modulus ratio                             & DFT         \\
            $P_C$               & Cauchy's pressure                                & DFT         \\
            $v_l$               & Longitudinal wave velocity                       & DFT         \\
            $v_t$               & Transverse wave velocity                         & DFT         \\
            $v_r$               & Velocity ratio                                   & DFT         \\
            $\langle v \rangle$ & Average wave velocity                            & DFT         \\
		$$ & $$ & $$ \\
\hline
\hline
		$$ & $$ & $$ \\
            Symbol                        & Descriptor (D-V)   &    Estimation approach            \\ 
		$$ & $$ & $$ \\
\hline
\hline
		$$ & $$ & $$ \\
	    $\gamma$                      & Lindemann ratio                         & AIMD      \\
            $\Gamma$                      & Lowest-energy optical phonon mode       & DFT       \\	
	    $\langle \omega \rangle$      & Mean frequency                          & AIMD      \\
            $\langle \omega_{30} \rangle$ & Mean frequency (cut-off at $30$~meV)    & AIMD       \\
            $E_{vib}$                     & Vibrational phonon energy               & AIMD       \\
            $C_v$                         & Constant volume heat capacity           & AIMD       \\
            $\theta_D$                    & Debye temperature                       & AIMD       \\
	    $F_{vib}$                     & Vibrational Helmholtz free energy       & AIMD       \\
            $S_{vib}$                     & Vibrational entropy                     & AIMD       \\
            $D$                           & Diffusion coefficient                   & AIMD       \\
   	    $msd$                         & Mean-squared displacement               & AIMD       \\
            $$ & $$ & $$ \\
\hline
\hline
		$$ & $$ & $$ \\
	    Symbol                                & Descriptor (S-C)                 &    Estimation approach         \\ 
		$$ & $$ & $$ \\
\hline
\hline
		$$ & $$ & $$ \\ 
	    $Z_N$               & Nominal charge                                     & Formula     \\
            $Z_B$               & Born effective charge                              & DFT         \\
            $\epsilon$          & Ion-clamped macroscopic dielectric constant        & DFT         \\
    	    $M$                 & Mobile ion atomic mass                             & Formula     \\
	    $\alpha_I$          & Mobile ion polarizability                          & DFT         \\
  	    $\alpha_C$          & Crystal polarizability                             & DFT         \\
    	    $Stc$               & Stoichiometry                                      & Formula     \\
  	    $Sym$               & Crystal symmetry                                   & DFT         \\
	    $a_m$               & Minimal lattice constant                           & DFT         \\
	    $n$                                   & Number of formula units          & DFT         \\
	    $\Omega$                              & Volume per formula unit          & DFT         \\
	    $\langle a b c \rangle$               & Standard deviation of lattice constants  & DFT         \\
	    $\langle \alpha \beta \gamma \rangle$ & Standard deviation of lattice angles     & DFT         \\
	    $SO$                                  & Number of crystal symmetry operations    & DFT         \\
	    $N_{nn}$                              & Number of nearest neighbors              & DFT         \\
	    $d_{nn}$                              & Nearest neighbors distance       & DFT         \\
	    $E_g$                                 & Band gap                         & DFT         \\
	    $E_{vac}$                             & Vacancy energy of the mobile ion         & DFT         \\
		$$ & $$ & $$ \\
\hline
\hline
    \end{tabular}
    \end{tabular}
    \caption{Analyzed SSE descriptors and their abbreviations. The materials descriptors were classifed into the categories (1)~``mechanical 
       	     and elastic'' (M-E), (2)~``diffusive and vibrational'' (D-V) and (3)~``structural and compositional'' (S-C). The method of calculation 
	     of each descriptor, either zero-temperature (DFT) or finite-temperature (AIMD) simulations, is indicated in the third column.
	     Some descriptors were directly deduced from the compounds formula, indicated as ``Formula'' in the table.}
\end{table*}

{\bf SSE descriptors abbreviations.}~To analyze the similarities and dissimilarities between fast-ion conductors a great variety
of different physical descriptors were estimated for each SSE, which are summarized in Table~I. The descriptors are generally
classified according to the quality they refer to, in particular: ``mechanical-elastic'' (M-E), ``diffusive-vibrational'' (D-V) and
``structural-compositional'' (S-C). It may be noted that most D-V descriptors like the mean phonon frequency (both with and without
cut-off), harmonic phonon energy, constant-volume heat capacity, Helmholtz free energy and entropy, were calculated for the materials
as a whole (i.e., considering both diffusive and non-diffusive ions) and also exclusively considering either the non-diffusive (denoted
as ``nd'' in the figures) or diffusive atoms (denoted as ``d'' in the figures). The total number of descriptors considered in this
work is equal to $54$. The descriptors estimated from AIMD (DFT) simulations were obtained at $T = 500 \pm 100$~K ($T = 0$~K).
\\

\section*{Data availability}
The data that support the findings of this study are available upon reasonable request from the authors C.L. and C.C. and 
the URL: https://superionic.upc.edu/

\section*{Acknowledgements}
We acknowledge financial support from the MCIN/AEI/10.13039/501100011033 under Grant No.
PID2020-119777GB-I00, the ``Ram\'on y Cajal'' fellowship RYC2018-024947-I, the Severo Ochoa 
Centres of Excellence Program (CEX2019-000917-S), the Generalitat de Catalunya under 
Grant No.2017SGR1506, and the CSIC under the ``JAE Intro SOMdM 2021'' grant program. 
\\

\section*{Author contributions}
C.C. conceived the study and planned the research, which was discussed in depth with the rest of co-authors. 
C.C. and R.R. performed and analyzed the first-principles calculations. C.L. carried out the data analysis 
of the generated DFT-AIMD database as well as the training of the SSE machine learning models. A.E. created 
the website that gives access to the DFT-AIMD database. The manuscript was written by C.C. with substantial 
input from the rest of co-authors.
\\

\section*{Additional information}
Supplementary information is available in the online version of the paper.
\\

\section*{Competing financial interests}
The authors declare no competing financial interests.

\end{document}